\documentclass[12pt,preprint]{aastex}

\tighten
\def\lsim{\lower.5ex\hbox{$\; \buildrel < \over \sim \;$}}
\def\gsim{\lower.5ex\hbox{$\; \buildrel > \over \sim \;$}}
\def\lax {\ifmmode{_<\atop^{\sim}}\else{${_<\atop^{\sim}}$}\fi}
\def\gax {\ifmmode{_>\atop^{\sim}}\else{${_>\atop^{\sim}}$}\fi}
\def\etal{{\it et al.\/} }
\def\gtorder{\mathrel{\raise.3ex\hbox{$>$}\mkern-14mu
\lower0.6ex\hbox{$\sim$}}}
\def\ltorder{\mathrel{\raise.3ex\hbox{$<$}\mkern-14mu
\lower0.6ex\hbox{$\sim$}}}

\def\pmb#1{\setbox0=\hbox{#1}%
\kern-0.015em\copy0\kern-\wd0
\kern0.03em\copy0\kern-\wd0
\kern-0.015em\raise0.0433em\box0 }

\begin{document}

\title{B-field Determination from Magnetoacoustic Oscillations in kHz QPO Neutron Star
 Binaries: Theory and Observations}

\author{L. G. Titarchuk}
\affil{George Mason University/CEOSR, Fairfax VA}
\email{lev@xip.nrl.navy.mil}

\author{C. F. Bradshaw}
\affil{School of Computational Sciences, George Mason University,
Fairfax VA} \email{cbradshaw@tstag.com}

\author{ K.S. Wood}
\affil{Naval Research Laboratory, Washington DC}
\email{wood@ssd0.nrl.navy.mil}
\vskip 0.5 truecm

%c\font\rom=cmr10
%c\centerline{\rom Accepted, Astrophys. J. }

\begin{abstract}
We present a method for determining the B-field around neutron stars
based on observed kHz and viscous QPO frequencies used in combination with the best-fit
optical depth and temperature of a Comptonization model.
In the framework of the transition layer QPO model,
we analyze magnetoacoustic wave formation in the layer between a
neutron star surface and the inner edge of a Keplerian disk.
We derive formulas for the magnetoacoustic wave frequencies for
different regimes of radial transition layer oscillations. We
demonstrate that our model can use the QPO as a new kind of probe to determine
the magnetic field strengths for 4U 1728-42, GX 340+0, and Sco X-1 in the zone where the QPOs
occur. Observations indicate that the dependence of the viscous frequency
on the Keplerian frequency is closely related to the inferred
dependence of the magnetoacoustic wave frequency on the Keplerian
frequency for a dipole magnetic field. The magnetoacoustic wave dependence
is based on a single parameter, the magnetic moment of the star as estimated from the field
strength in the transition layer. The best-fit magnetic moment parameter is about $(0.5-1)\times10^{25}$
G cm$^3$ for all studied sources. From  observational data, the
magnetic fields within distances less 20 km from neutron star for all three sources
 are strongly constrained to be
dipole fields with the strengths $10^{7-8}$ G on the neutron star surface.
\end{abstract}

%\keywords{X-rays: bursts-stars --- accretion disks, LMXB, X-ray
%binary --stars: neutron ---stars:
%individual: 4U 1728-34,  GX 340+0, Sco X-1}


\section{Introduction}

%The most reliable estimates of magnetic field strengths for neutron
%stars (NS) are from radio pulsars, with field strengths derived from spin down
%typically of order $\sim 10^{12}$ G
%(Bhattacharya, \& Srinivasan 1995).  Millisecond
%pulsars are an exception to this clustering, with magnetic fields
%less than $10^9$G. 
Strong magnetic fields of neutron stars, typically of order $\sim 10^{12}$ G (Bhattacharya, \& Srinivasan 1995)
are considered
crucial to the radio pulsar activity but equilibrium arguments concerning
NS magnetic field pressure and ram pressure balance lead to magnetic
field strength estimates
for heavily accreting NS sources (for example X-ray bursters) of $\ltorder10^9$ G.
In these sources no persistent pulsations have been found despite careful searches
(Wood et al. 1991; Vaughan et al. 1994).
Different scenarios of magnetic field evolution lead to different
estimates of magnetic field strength and decay. Thus, it is desirable
to find a direct way of measuring or estimating the neutron star's magnetic
field in these binary systems. Any model used to make such estimates should have minimum assumptions about 
observed data and model relationships to produce accurate
results. {\it Our model, described herein, has only one parameter that can be adjusted
for agreement, the magnetic moment.
 All other parameters are strongly constrained by observations.}
It develops a new probe of the magnetic field strength and geometry, based on radial
magnetoacoustic oscillations.

Titarchuk, Lapidus \& Muslimov (1998), hereafter TLM98,
considered the possibility of dynamical adjustments of a Keplerian
disk to the innermost sub-Keplerian boundary conditions to explain
most observed QPOs in bright low mass X-ray binaries (LMXBs).
They concluded that an isothermal transition sub-Keplerian layer
between the NS surface and its last Keplerian orbit forms as a result
of this adjustment. The TLM98 model is general treatment  applicable to both NS and
black hole (BH) systems. The primary problem in both NS and BH systems is
understanding how the flow changes from pure Keplerian to sub-Keplerian as a radius
decreases to small values. The TLM98 authors suggested
that the discontinuities and abrupt transitions in their solution result
from derivatives of quantities such as angular velocities (weak shocks).
Subsequently, Titarchuk \& Osherovich (1999 hereafter, TO99) identified
the transition layer (TL) model's corresponding TL
viscous frequency $\nu_V$ and diffusion or break frequency
$\nu_{b}$ for 4U 1728-34, and predicted values for
$\nu_b$ related to diffusion in the boundary layer.
%Then, Titarchuk, Osherovich \& Kuznetsov (1999 hereafter TOK99)
%confirmed, with observations of a number of QPO sources (Wijnands et
%al. 1999),  a relationship, valid for both NS and BH candidate sources, of
%$\nu_b\propto \nu_V^{1.6}$.
TLM98 argued that a shock should occur where the Keplerian disk adjusts
to  sub-Keplerian flow. They interpreted the low frequencies as acoustic oscillations
and expected them to provide reliable estimates of radial
velocities that would be very close to the acoustic velocities.

Sco X-1's derived acoustic velocities (Titarchuk et al. 2001 hereafter
TBGF) translated to a TL plasma temperature, assuming only thermal motion, 
of about 19 keV. However photon spectral fitting by TBGF yielded a TL temperature of 2.8 keV.
TBGF attributed this discrepancy to a dominant contribution to
the plasma matter energy density from the magnetic field. 
%TBGF estimated
%the magnetic field strength to be $B < 0.7\times10^6$G in the TL
%of Sco X-1 at about 1-1.5 NS radii from the surface.
{\it The viscous QPOs provide information identifiable with a
magnetoacoustic (MA) oscillation effect that can be modeled}.

In this {\it Letter}, we develop  a general
formalism for MA radial oscillations in the
TL. In \S 2, we describe the data sources used to test our
MA oscillation model. The formulation of the problem and derivation of
MA frequencies are described in \S 3. We show the results of the
MA oscillation interpretation for the TO99 QPO viscous frequencies
and  the best-fit normalization parameters of magnetic field strengths
for a number of QPO sources  are present in \S 4.
Our summary and conclusions follow in \S 5.

\section{Observations}
The model can only be applied to QPOs in which the full TLM has been worked out and in which
the QPOs identified with oscillations of a bounded medium have been seen.
All observations\footnote{This research has made use of data obtained
through the High Energy Astrophysics Science Archive Research Center
Online Service, provided by the NASA/Goddard Space Flight Center.}
referenced herein were made with the Rossi X-ray Timing Explorer's (RXTE).
Sco X-1 data were  obtained during August 3 and 22, 1997, February 27
and 28, 1998, and June 10-13, 1999.
Data were extracted from research archives for GX 340+0 and 4U 1728-34 for
observations on November 14, 1998 and February 28, 1999, respectively.
In addition, we use the results from observations of 4U 1728-32 between
February 15-March 1, 1996 by Ford \& van der Klis (1998)

\section{Magnetoacoustic Oscillations in the Transition Layer}

In the references cited above, the treatment of a magnetic field in a fluid disk
has specified neither the multipole order that characterizes the field
nor the strength of the field. 
%Yet, the
%invariance of $\delta$ (Osherovich \& Titarchuk 1999) in NS systems
%can be viewed as evidence that the central object is
%magnetized because it behaves like an object of finite
%radius with a symmetry-breaking feature. 
%The feature, which is the
%magnetic field, channels plasma flow nearly
%perpendicular to the disk along an angle that departs from being
%perpendicular by a small invariant amount.
%Osherovich, Tzur \& Gliner (1984) first showed that the quadrupole term
%introduces north-south asymmetry in the makeup of the magnetic atmosphere
%surrounding a gravitating object. Recently, this asymmetry phenomenon has been fully
%confirmed for the solar corona (e.g., Gibson et al. 1999).

We derive the frequency of the QPO associated with MA oscillations
and the correlation of the MA frequency with the Keplerian
frequency $\nu_{\rm K}$. The MA frequency is derived as the eigenfrequency
of the boundary-value problem resulting from a MHD treatment of the
interaction of the disk with the magnetic field.
The problem is solved for two limiting boundary conditions encompassing
realistic possibilities. The solution yields a velocity
identified as a mixture of the sound speed and the Alfv\'en velocity,
becoming either at the appropriate limits.  Our treatment does not specify
how the eigenfrequency is excited or damped. However, it makes clear that the QPO
is a readily stimulated resonant frequency.

In order to derive the MA oscillation frequencies in the
isothermal transition layer, we consider layer perturbations such that the
density $\rho=\rho_0+\rho_1$, the velocity ${\bf v=v_0+v_1}$ and ${\bf
B=B_0+B_1}$. By combining the continuity equation, equation of motion, and an ideal
gas equation, we  obtain an equation for MA oscillations (Alfv\'en 1942)
\begin{equation}
{{\partial^2 {\bf v_1}}\over{\partial
t^2}}-s^2\nabla(\nabla\cdot{\bf v_1})
+{\bf v_{\rm A}}\times\nabla\times[\nabla\times({\bf
v_1\times v_{\rm A}})]=F({\bf v_0, \partial v_0/\partial r, B_0}, \rho_0)
\label{eq:alven}
\end{equation}
where $s^2=(\partial P/\partial \rho)$ is the square of the sound velocity
 (assumed constant with respect to radius)
and ${\bf v_{\rm A}}={\bf B_0}(4\pi\rho_0)^{-1/2}$ is the
Alfv\'en velocity.
To determine the eigenfrequencies of
MA oscillations in the TL, we set equation
(\ref{eq:alven}) to zero and introduce a cylindrical coordinate system in
the TL with a radial axis in the disk plane, z-axis perpendicular to
the plane, and azimuthal axis along the circle in the disk plane
[see  Fig. 1 in  Titarchuk, Osherovich \& Kuznetsov (1999), hereafter TOK,
  for the transition layer geometry].

The magnetic field acts along a  direction in space that is the same throughout the TL,
intersecting the disk everywhere at the same angle $\delta$.
Practical ranges for $\delta$ are
from a few to $\sim 15$ degrees (Titarchuk \& Osherovich 2001).
The model used for all three sources (see \S 4 )
assumes that $\delta=0$ exactly, which reinstates
axial symmetry and makes the problem mathematically tractable.
The simplified problem is spatially one dimensional in the radial direction, 
and oscillations represented by the eigenfunctions are radial fluid movements.

We consider the radial oscillations in the TL
for which we should assume that ${\bf B_0}=B_0{\bf e_z}$, i.e., the mean
magnetic field strength ${\rm B_0}$ is perpendicular to the disk and
${\bf v_1}= v_1{\bf e_{r}}$ is along the radial direction in the disk.
We wish to explore how solutions for flow in the TL are affected by boundary conditions 
and B field geometry, for acoustic and  magnetic  oscillations.
Under these assumptions, equation (\ref{eq:alven}) can be rewritten
as:
\begin{equation}
{{\partial^2 v_1}\over{\partial t^2}}-s^2{{\partial}\over{\partial
r}}\left({1\over r}{{\partial r v_1}\over{\partial r}}\right)- v_{\rm A}
{{\partial}\over{\partial r}}\left({1\over r}
{{\partial r v_{\rm A}v_1}\over{\partial r}}\right)=0.
\label{eq:new_alven}
\end{equation}
Combining this equation with the two boundary conditions of the
TL, the neutron star surface $r=R_{in}$
and the outer boundary, $r=R_{out}$, defined by the Keplerian QPO
frequency, allows us to analyze the eigenfrequencies of the MA
oscillations in the TL under both the stiff
$v_1(R_{in})=v_1(R_{out})=0$
 and the free $\partial v_1/\partial r(R_{in})=\partial v_1/\partial r(R_{out})=0$ 
 boundary conditions.

In the two extreme cases, either $v_{\rm A}\ll s$ or  $v_{\rm A}\gg s$
equation (\ref{eq:new_alven}) can be examined
analytically under the appropriate boundary conditions.  Introducing a
new function $y=rv_1$ in the case of $v_{\rm A}\ll s$ and, $y=rv_{\rm
A}v_1$,  in the case of $v_{\rm A}\gg s$, we can write equation
(\ref{eq:new_alven}) in the form
\begin{equation}
{{\partial^2 y}\over{\partial
t^2}}=4Ax^{(2-\alpha)/2}{{\partial^2y}\over{\partial x^2}}
\label{eq:partial_alven}
\end{equation}
where  $x=r^2$ is a new variable, $\alpha =0$ and $A=A_s=s^2$ for the
pure acoustic case, and $\alpha \geq 6$ and
$A=A_{m}=B_{\ast}^2r_{\ast}^{\alpha}/4\pi\rho$
for the magnetic case. The index $\alpha$ is related to the multipole
magnetic field,  $v_{\rm A}^2=A_{m}/r^{\alpha}$ ($\alpha=6, 8, 10$ are for
the dipole, quadrupole, and octopole, respectively).
 $B_{\ast}$ is the magnetic field strength at a radius
$R_{in}<r_{\ast}<R_{out}$, and $\rho$ is the TL mean density.

First, we need to find a solution of the eigenvalue problem for
equation (\ref{eq:partial_alven}) under the stiff boundary
conditions by obtaining a particular nontrivial solution using the
separation variables method, such that
\begin{equation}
y(x,t)=X_{\lambda}(x)T_{\lambda}(t)
\end{equation}
where $X_{\lambda} (x)$ is an eigenfunction satisfying
\begin{equation}
X^{\prime\prime}+{{x^{(\alpha-2)/2}}\over{4A}}\lambda^2 X=0
\end{equation}
and boundary conditions $X(x_{in})=X(x_{out})=0$.
The solution of these equations can be expressed through the Bessel
$Z$ functions
\begin{equation}
X_{\lambda}=x^{1/2}
Z_{2/(\alpha+2)}\left[{{2}\over{(\alpha+2)}}{{\lambda}\over{A^{1/2}}}
x^{(\alpha+2)/4}\right].
\end{equation}
$X_{\lambda}$ is a linear superposition of the Bessel
$J_{2/(\alpha+2)}$ and $J_{-2/(\alpha+2)}$ functions:
\begin{equation}
X_{\lambda}=x^{1/2}
[J_{2/(\alpha+2)}(z_{\lambda})~+~CJ_{-2/(\alpha+2)}(z_{\lambda})]
\label{eq:besselj}
\end{equation}
where
\begin{equation}
z_{\lambda}={{2}\over{(\alpha+2)}}{{\lambda}\over{A^{1/2}}}x^{(\alpha+2)/4}.
\label{eq:besselz}
\end{equation}
Wood et al. (2001) demonstrated that an asymptotic form
of the Bessel function for $z_{\lambda}\gax 1$  can be used to find
all eigenvalues  including the first one:
\begin{equation}
J_{2/(\alpha+2)}(z_{\lambda})=\left({2\over{\pi
z_{\lambda}}}\right)^{1/2}
\cos[z_{\lambda}-(\alpha+6)\pi/4(\alpha+2)]
\end{equation}
and
\begin{equation}
J_{-2/(\alpha+2)}(z_{\lambda})=\left({2\over{\pi
z_{\lambda}}}\right)^{1/2}
\cos[z_{\lambda}-(\alpha-2)\pi/4(\alpha+2)].
\end{equation}

Using these asymptotic forms, the boundary conditions, and equations
(\ref{eq:besselj} \& \ref{eq:besselz}),
the eigenvalues are determined by
\begin{equation}
\cos(z_{out}-\varphi_1)\cos(z_{in}-\varphi_2)-\cos(z_{out}-\varphi_2)\cos(z_
{in}-\varphi_1)=0,
\label{eq:eigen}
\end{equation}
where $\varphi_1=(\alpha-2)\pi/4(\alpha+2)$ and
$\varphi_2=(\alpha+6)\pi/4(\alpha+2)$.
Equation (\ref{eq:eigen}) can be reduced to
$\sin(z_{out}-z_{in})=0$ or $\beta=z_{out}-z_{in}=n\pi;~~n=1,2,3...$,
with a solution
\begin{equation}
\lambda_n={{\alpha+2}\over{2}}{{n\pi
A^{1/2}}\over{r_{out}^{(\alpha+2)/2} -r_{in}^{(\alpha+2)/2}}}.
\end{equation}
The time dependence, $T(t)$ is found from the equation
$T^{\prime\prime}+\lambda_n^2T=0$.
Thus, $T(t)=\sin(\lambda_n t+\phi_0)=\sin(2\pi\nu_n t+\phi_0)$ where
$\phi_0$ is an initial phase. For the acoustic case ($A=A_s=s$ and $\alpha=0$), 
the main eigenvalue
$\lambda_1=\pi /(r_{out}-r_{in})= \pi/L$ and
$\nu_s=\nu_1=\lambda_1/2\pi=s/2L$,
where $L$ is the TL radial size.

For the magnetic case ($\alpha\geq6$)
\begin{equation}
\nu_{M}=\nu_1=
{{(\alpha+2)B_{\ast}}\over{4(4\pi\rho)^{1/2}[r_{out}-r_{in}(r_{in}/r_{out})^
{\alpha/2}]}}\left({{r_{\ast}}\over{r_{out}}}\right)^{\alpha/2}={{(\alpha+2)
v_{\rm A}(r_{out})}\over{4[r_{out}-r_{in}(r_{in}/r_{out})^{\alpha/2}]}}.
\end{equation}
For the free boundaries, the boundary conditions for $y$ are
\begin{equation}
{{\partial y}\over{\partial x}} +{{(\alpha-2)}\over{4}}{y\over
x}=0~~~~{\rm at}~~ x=x_{in}~~ {\rm and}~~x=x_{out},
\end{equation}
from which, the transcendental equation for determining the eigenvalue is
\begin{equation}
\tan\beta=-{{2(\alpha-2)\beta}\over{(\alpha+2)\{\eta\beta^2/
(\eta-1)^2+[(\alpha-2)/(\alpha+2)]^2\}}},
\label{eq:eigen_tan}
\end{equation}
where
$\beta=z_{out}-z_{in}$ and
$\eta=z_{out}/z_{in}=(r_{out}/r_{in})^{(\alpha+2)/2}$.

Analysis of equation (\ref{eq:eigen_tan}) shows that for
the values most relevant to observations ($\eta\gax 2$),
\begin{equation}
\beta_{M}\approx
{{\pi}\over2}+{2\over{\pi}}[(\pi/2)^2\eta/(\eta-1)^2+1/4]
\end{equation}
for $\alpha=6$ and $\beta_{s}\approx \{1.5/[1+1.5\eta/(\eta-1)^2]\}^{1/2}$
for $\alpha=0$.

The free-boundary conditions will therefore introduce additional factors appearing in
formulas for $\nu_1$ between 0.5 and 1 for the magnetic case and  
$1/\pi$ for the acoustic case.
From this, we conclude that the free-boundary conditions can decrease substantially
the magnetic oscillation frequency and must 
decrease the acoustic frequency by at least a factor of 3 (see Eqs. 17-18 below).

\section{Magnetic Field Strength Determination and the Magnetoacoustic
Oscillation Frequency}

We can construct an approximate formula for the MA frequency
$\nu_{MA}$ using the asymptotic forms for the two extreme (acoustic
and magnetic) cases
\begin{equation}
\nu_{MA}\approx (\nu_{s}^2+\nu_M^2)^{1/2}=
\{s^2/4(r_{out}-r_{in})^2+[(\alpha+2)/4]^2v_{\rm
A}^2(r_{out})/[r_{out}-r_{in}(r_{in}/r_{out})^{\alpha/2}]^2\}^{1/2}.
\label{eq:ma_freq}
\end{equation}
The approximate relation in equation (\ref{eq:ma_freq}) is exact if
the Alf\'ven velocity is assumed to be constant
through the TL.  The difference, by a factor
$(\alpha+2)^2/4$, in
coefficients for $s^2$ and $v_{\rm A}^2$ is due to different behaviors
of $s$ and  $v_{\rm A}$ as a function of $r$ ($s=const$ and $v_{\rm
A}\propto r^{-\alpha/2}$).

Under the free boundary, equation (\ref{eq:ma_freq}) is modified to
\begin{equation}
\nu_{MA}\approx \{(\beta_s/\pi)^2s^2/4(r_{out}-r_{in})^2+
(\beta_{M}/\pi)^2[(\alpha+2)/4]^2v_{\rm
A}^2(r_{out})/[r_{out}-r_{in}(r_{in}/r_{out})^{\alpha/2}]^2\}^{1/2}
\label{eq:ma_free}
\end{equation}
and provides the dependence of $\nu_{MA}$ on $\nu_{\rm K}$.  This
dependence is due to the explicit dependence of $r_{out}$ on the
Keplerian frequency
$\nu_{\rm K}$:
\begin{equation}
r_{out}=(GM/2\pi\nu_{\rm K}^2)^{1/3},
\label{eq:rout}
\end{equation}
where $G$ is the gravitational constant and $M$ is the NS mass. 
To determine $s$, we define $s=(kT/m_p)^{1/2}$ as the sound speed for the
isothermal TL where $kT$ is the best-fit value of the Comptonization temperature.
To determine $v_{M}=B(r_{out})/(4\pi \rho_0)^{1/2}$, we express
$\rho=\tau m_p/(\sigma_T L)$ with $\tau$ as the best-fit optical depth
parameter. These are not adjustable parameters for
determining magnetic field strengths.

{\it Thus, we obtain $\nu_{MA}$ as a function of only one adjustable parameter,
the magnetic-field strength at a TL radius
$r_{\ast}$}. We can  replace this parameter by the multipole moment
$B(r_{\ast})r_{\ast}^{\alpha/2}$. In the dipole case $\alpha=6$, the multipole moment is the 
dipole moment $\mu$ (in G cm$^{3}$).
 The procedure for determining the magnetic field strength in the
TL may realized if there is an observable dependence
between the viscous and Keplerian frequencies.

The spectra of 4U 1728-34, GX 340+0, and Sco X-1 were fitted by
the Comptonization model of Sunyaev \& Titarchuk (1980) from
8-25 keV, to obtain a plasma temperature
(kT) and Thompson optical depth ($\tau$).
Figure 1 shows the results from the model and the fitted data
for 4U 1728-34, GX 340+0, and Sco X-1 (Ford \& van der Klis 1998).

We found that a dipole geometry $(\alpha=6)$ gives a good fit and higher
poles are excluded by observed data.
It is also evident from equations (\ref{eq:ma_freq}-\ref{eq:rout}) that
the MA frequencies are
well described by a power law dependence on $\nu_{\rm K}$ when
$\nu_M>\nu_s$, i.e.,
$\nu_{MA}\approx \nu_{M}\propto \nu_{\rm K}^{(\alpha+2)/3}$. The  power law index in the
dipole case
is $(\alpha+2)/3=8/3$.
The MA frequency dependence on $\nu_{\rm K}$  is too steep
for the quadrupole and the octopole cases  [where $(\alpha+2)/3=10/3$ and
4, respectively] to be supported by the data.
The fits of our MA model to the data provide strong constraints
for the boundary condition types where the stiff conditions are ruled out
for 1728-34 and the free conditions for GX 340+0. The fit qualities are practically
 insensitive for  $M =1.2\pm 0.2 M_{\odot}$. The best-fit mass  parameter   is around 1.2.

The best-fit parameters for 4U 1728-34 (Compton values $kT=8.6 \pm 2$ keV,
$\tau = 5.8 \pm 0.5$; $\chi^2/dof = 19.8/35$)  
%are obtained 
%as the best-fit parameters of the Comptonization model to
%the X-ray spectrum ) 
are $B_{\ast}=(0.86\pm0.03)\times10^6$ G,
%$m=M/M_{\odot}=1.1\pm 0.2$ 
and $r_{\ast}=1.69\times10^6$ cm,
corresponding to a magnetic moment $\mu=4.2\times10^{24}$ G cm$^3$.
We fixed the NS radius, $r_{in}$ to be (as a canonical value) three Schwarszchild radii $3~r_{\rm S}$. An
extrapolation of the magnetic field from $r_{\ast}$ towards the NS radius
$r_{in}$ gives us $B_{NS}=4.5\times10^6$ G and $B_{NS}=1.3\times10^7$ G
for the dipole and octopole  fields, respectively.

The best-fit parameters for GX 340+0 (Compton values $kT=2.87 \pm 0.02$
keV, $\tau = 11.2 \pm 0.2$;
$\chi^2/dof = 49.8/35$) are $B_{\ast}=(0.5\pm0.05)\times10^6$ G
and 
%$m=1.2\pm 0.2$ and  
$r_{\ast}=2.7\times10^6$ cm,
corresponding to a magnetic moment $\mu=1\times10^{25}$ G cm$^3$. An
extrapolation of the magnetic field
from $r_{\ast}$ towards the NS radius $r_{in}$ yields
$B_{NS}=8.3\times10^6$ G and $B_{NS}=5.3\times10^7$ G   for the dipole
and octopole fields, respectively.

%Our free-boundary condition provided the best solution to meet
%the data constraints for  Sco X-1. 
The best-fit parameters for Sco X-1
(Compton values $kT=2.87 \pm 0.03$ keV, $\tau = 11.3 \pm 0.3$; $\chi^2/dof
=9.9/33$) are $B_{\ast}=(1.\pm0.05)\times10^6$ G,  
%$m=1.2\pm 0.2$  
and 
$r_{\ast}=2.12\times10^6$ cm,
corresponding to a magnetic moment $\mu=9.5\times10^{24}$ G cm$^3$ in
the TL. An extrapolation of the magnetic field
from $r_{\ast}$ towards the NS radius $r_{in}$ gives us
$B_{NS}=8.1\times10^6$ G and $B_{NS}=3.3\times10^7$ G  for the dipole
and octopole fields, respectively.

These derived magnetic-field strengths are
for the TL only and can be smaller than the true
magnetic field outside of the layer because of its finite conductivity.
The NS magnetic-field strengths  can be  increased by factors of 1.7 and 2.5
for the dipole and octopole extrapolation respectively if we assume that the NS radius 
is around $2.5~r_{\rm S}$ instead of  $3~r_{\rm S}$. 
% In fact, Haberl \& Titarchuk (1995) have
%already shown such a NS mass-radius relation for 4U 1820-30 and recently they obtained a similar 
%relation for 4U 1728-34 (Titarchuk \& Haberl 2001). 
   
%The model, in its simplest form, provides  a surprisingly good fit to the variation
%seen in all three sources without difficulty and without further parametrization beyond
%fitting the magnetic dipole moment. Values obtained in all three sources are
%reasonable. 
The  magnetic field strength  can be roughly checked using an
equipartition estimate for which $B^2/8\pi=nkT$.
For example, Sco X-1's  TL density $\rho=2.7\times10^{-5}$ g cm$^{-3}$
is obtained using the best-fit optical depth $\tau=11.3$.
The TL size, $L=10^6$ cm,  is obtained using the observed
Keplerian frequency and best-fit temperature $kT= 2.9$ keV.
Thus, we obtain a magnetic field, $B=1.3\times10^6$ G, at radius
$r=2.3\times10^6$ cm.

\section{Summary and Conclusions}
Since the discovery of neuron stars in the 1960s magnetic fields have been estimated almost
exclusively from {\it rotational} effects, i.e. from periodic pulses with their 
periods $p$ and derivatives $\dot p$ [e.g. Bhattacharya \& Srinivasan (1995)]. The new idea here is that, when the field configuration is nearly 
axis-symmetric, there may arise circumstances in which the field is estimated from
radial oscillations (QPOs) of plasma interacting with magnetosphere.

%We have presented a model that goes beyond TO99 to treat the radial oscillations in the
%transition layer surrounding a neutron star.
The model preconditions are:
(1) the MA QPO must be detected observationally, (2) a Keplerian QPO
frequency must be observed and (3) the temperature and the optical
depth of the QPO emission region must be obtained from the photon spectra.
From  our model we derived tightly constrained dipole magnetic
fields of the TL for the bright LMXBs 4U 1728-34, GX 340+0,
and Sco X-1.  The observed low frequency correlation of the kHz QPOs
strongly confine the theoretical dependence of
$\nu_{MA}$ on $\nu_{\rm K}$, which is mostly determined by the
magnetic frequency dependence on $\nu_{\rm K}$
(see dashed lines in Figure 1) and very sensitive
to the pole order (dipole, quadrupole, {\it etc.}).
The correction to acoustic oscillations [the first term in formulas (17-18)]
only slightly changes
the theoretical shape of the low frequencies of $\nu_{\rm K}$.

%Our analysis further confirms that the low-noise QPO frequency
%is a radial-oscillation frequency of the TL as a whole and
%demonstrates that it is dictated by MA oscillations. In addition, the
%absolute normalizations of these frequencies are determined only by the TL
%magnetic field strength, temperature, density, and neutron star mass.

We re-fit the stellar magnetic dipole moment in each source to derive a range of
values from $(0.5-1)\times 10^{25}$ G cm$^{3}$. 
%This range is plausible
%in all sources. We confirm that three LMXBs, characterized presumably
%(at least 4U 1728-32) by spin rates of order 300-400 Hz, have derived
%dipole field strengths on the order of $10^{6-7}$ G in the
%TL and on the order of $10^{7-8}$ G on the NS surface $10^{7-8}$ G. 
%These values support an evolutionary tie
%between LMXB sources and millisecond radio pulsars.
The field estimate arrived at this manner is close to that may be evolutionary end points
 of LMXBs, namely milisecond pulsars,  which have  derived
values (from $\sqrt{p\dot p}$) of  $\sim 10^8$ G (Bhattacharya 1995). 

We wish to thank Sergey Kuznetsov, Eric Ford, Peter Jonker and Michael
van der Klis for the Sco X-1, 4U 1728-34 and GX 340+0 data files.
We are truly grateful to the referee, J. Imamura for his thorough analysis of
the presented manuscript.

\clearpage

\clearpage
\begin{figure}
\epsscale{0.7}
\plotone{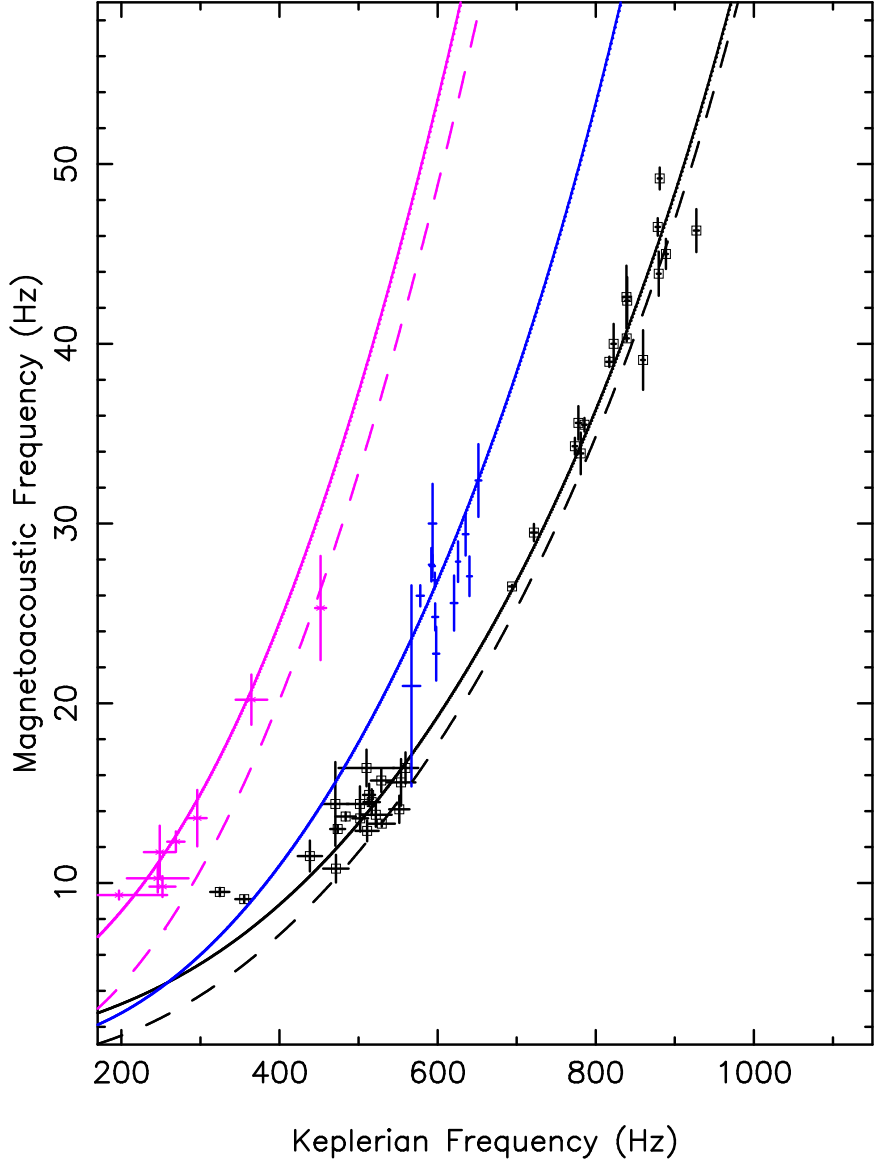}
\caption{MA frequency vs Keplerian frequency for 4U 1728-34
(black squares represent QPO frequencies from Ford \& van der Klis 1998,
blue, from TOK and magenta, from Jonker et al. 2000). 1-$\sigma$ errors bars indicate
statistical fitting errors
($\Delta\chi^2=1$).  The value for Q ($=\nu$/FWHM) approaches 1.0 for
viscous QPOs corresponding to Keplerian frequencies of 320-360 Hz (Ford \& van der Klis
1998), giving formal frequency errors significantly larger
than the plotted error bars. The solid lines are the theoretical best-fits for the MA
frequencies. The dashed lines are for the magnetic frequencies only (see text).}
\label{fig1}
\end{figure}
\clearpage

\end{document}